\documentclass[author, sn-nature]{sn-jnl}

\usepackage{graphicx}%
\usepackage{multirow}%
\usepackage{amsmath,amssymb,amsfonts}%
\usepackage{amsthm}%
\usepackage{mathrsfs}%
\usepackage[title]{appendix}%
\usepackage{xcolor}%
\usepackage{textcomp}%
\usepackage{manyfoot}%
\usepackage{booktabs}%
\usepackage{algorithm}%
\usepackage{algorithmicx}%
\usepackage{algpseudocode}%
\usepackage{listings}%

\unnumbered
\begin{document}

\title[Article Title]{\textbf{Memristive response and capacitive spiking in the aqueous ion transport through 2D nanopore arrays}}

\author[1,2]{\fnm{Yechan} \sur{Noh}}\email{yechan.noh@nist.gov}

\author*[1]{\fnm{Alex} \sur{Smolyanitsky}}\email{alex.smolyanitsky@nist.gov}

\affil[1]{\orgdiv{Applied Chemicals and Materials Division}, \orgname{National Institute of Standards and Technology}, \orgaddress{\city{Boulder}, \postcode{80305}, \state{Colorado}, \country{United States}}}

\affil[2]{\orgdiv{Department of Materials Science and Engineering}, \orgname{University of California, Berkeley}, \orgaddress{\city{Berkeley}, \postcode{94720}, \state{California}, \country{United States}}}

\maketitle
\vspace{-1.0cm} 
\textbf{In living organisms, information is processed in interconnected symphonies of ionic currents spiking through protein ion channels. As a result of dynamically switching their conductive states,
ion channels exhibit a variety of current-voltage nonlinearities and memory effects. Fueled by the promise of computing architectures entirely different from von Neumann, recent attempts to identify and harness similar phenomena in artificial nanofluidic environments focused on demonstrating analog circuit elements with memory. Here we explore aqueous ionic transport through two-dimensional (2D) membranes featuring arrays of ion-trapping crown-ether-like pores. We demonstrate that for aqueous salts featuring ions with different ion-pore binding affinities, memristive effects emerge through coupling between the time-delayed state of the system and its transport properties. We also demonstrate a nanopore array that behaves as a capacitor with a strain-tunable built-in barrier, yielding behaviors ranging from current spiking to ohmic response. By focusing on the illustrative underlying mechanisms, we demonstrate that realistically observable memory effects may be achieved in nanofluidic systems featuring crown-porous 2D membranes.}

Although the first memory-enabled resistor, or memristor, was formally proposed in 1971~\cite{1083337}, similar devices have been considered as interconnected building blocks in artificial neural networks since at least the 1960s~\cite{yang2013memristive}. The search for practical neuromorphic computing implementations has now evolved into a vast multidisciplinary field that ranges from materials science~\cite{yang2013memristive, Wang2018memcap, wang2020resistive, kumar2022dynamical, lanza2022memristive, christensen20222022, sangwan2020neuromorphic} and physics of dynamic systems~\cite{yang2013memristive, kumar2022dynamical, lanza2022memristive, sangwan2020neuromorphic} to information processing algorithms~\cite{jung2022crossbar, christensen20222022}. In the past five years, numerous nanofluidic devices featuring memory effects in their current-voltage response have been reported~\cite{bu2019nanofluidic, sheng2017transporting, zhang2019nanochannel, xiong2023neuromorphic, robin2023long, robin2021modeling, emmerich2023ionic}. The memory effects have been shown to arise from local time delays introduced by ion concentration polarization caused by high electric fields~\cite{bu2019nanofluidic}, diffusivity-limited dynamics~\cite{sheng2017transporting,zhang2019nanochannel,emmerich2023ionic}, the Wien effect~\cite{robin2023long, robin2021modeling}, and adsorption-desorption processes in confined electrolytes~\cite{robin2023long, emmerich2023ionic}. These mechanisms are of little surprise, resulting from the fundamental ubiquity of local time delays built into the dynamic response of essentially any realistic system, as predicted by the Kubo response theory~\cite{kubo1957statistical} and conceptually detailed by Di Ventra and Pershin specifically for memory-featuring versions of the basic circuit elements such as resistors, capacitors, and inductors~\cite{di2013physical}. In the context of nanofluidic systems, time delays arising from mechanisms that can be broadly classified as diffusion-limited and adsorption-desorption processes are required to build a realistic circuit element featuring memory. At the same time, in systems involving water as the only solvent, remembered states tend to dissipate rapidly, often within picoseconds. Therefore, reducing memory volatility is a fundamental challenge for harnessing memory effects in aqueous environments. Toward this goal, a clear understanding of the mechanisms that yield distinct and measurable memory effects in nanofluidic systems is critical. 
\newline
\newline
In this work, we use all-atom molecular dynamics (MD) simulations to describe highly illustrative memory and spiking phenomena in dynamically biased nanofluidic systems that do not rely on high-viscosity solvents, conical pore geometries, or 2D confinement of aqueous electrolytes. We first describe memristive transport through arrays of graphene-embedded crown pores in the presence of aqueous salt mixtures and explain how dynamic sieving of two salts, neither of which individually yields memory effects, can be a memristor through a straightforward coupling between the state of the system and its transport properties.  In addition, we demonstrate ion current spiking dynamics in the case of a hexagonal boron nitride (hBN) monolayer featuring an array of triangular nitrogen-terminated crown-like pores. Such a system is shown to be essentially a capacitive element with a built-in chemical barrier. We finally demonstrate that this barrier can be modulated by a tensile strain applied to the membrane, causing a transition in dynamic transport response from spiking to the tell-tale behavior of an RC-circuit.
\begin{figure}[ht!]
\centering
\includegraphics[width=1.0\linewidth]{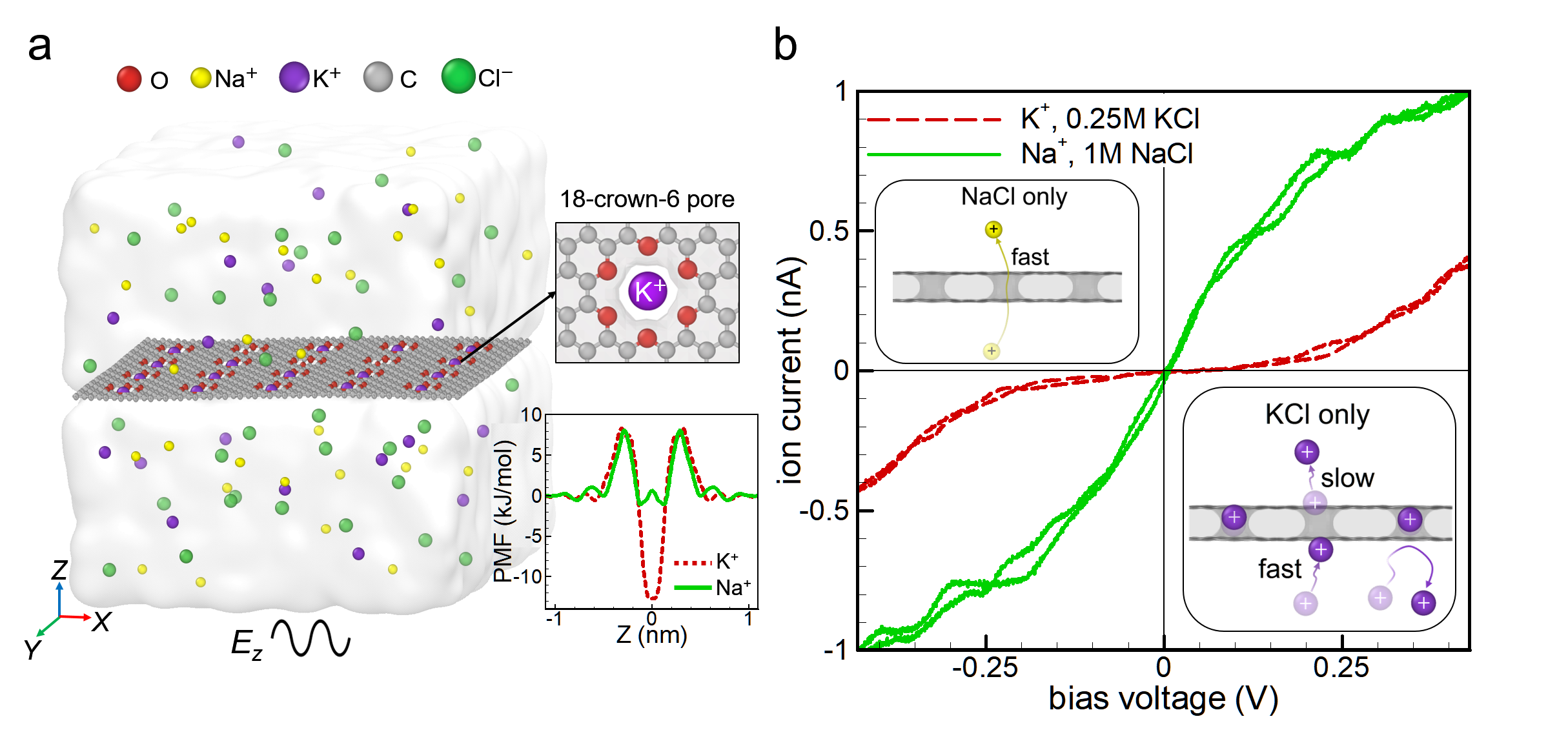}
  \vspace{-0.5cm} 
 \includegraphics[width=1.0\linewidth]{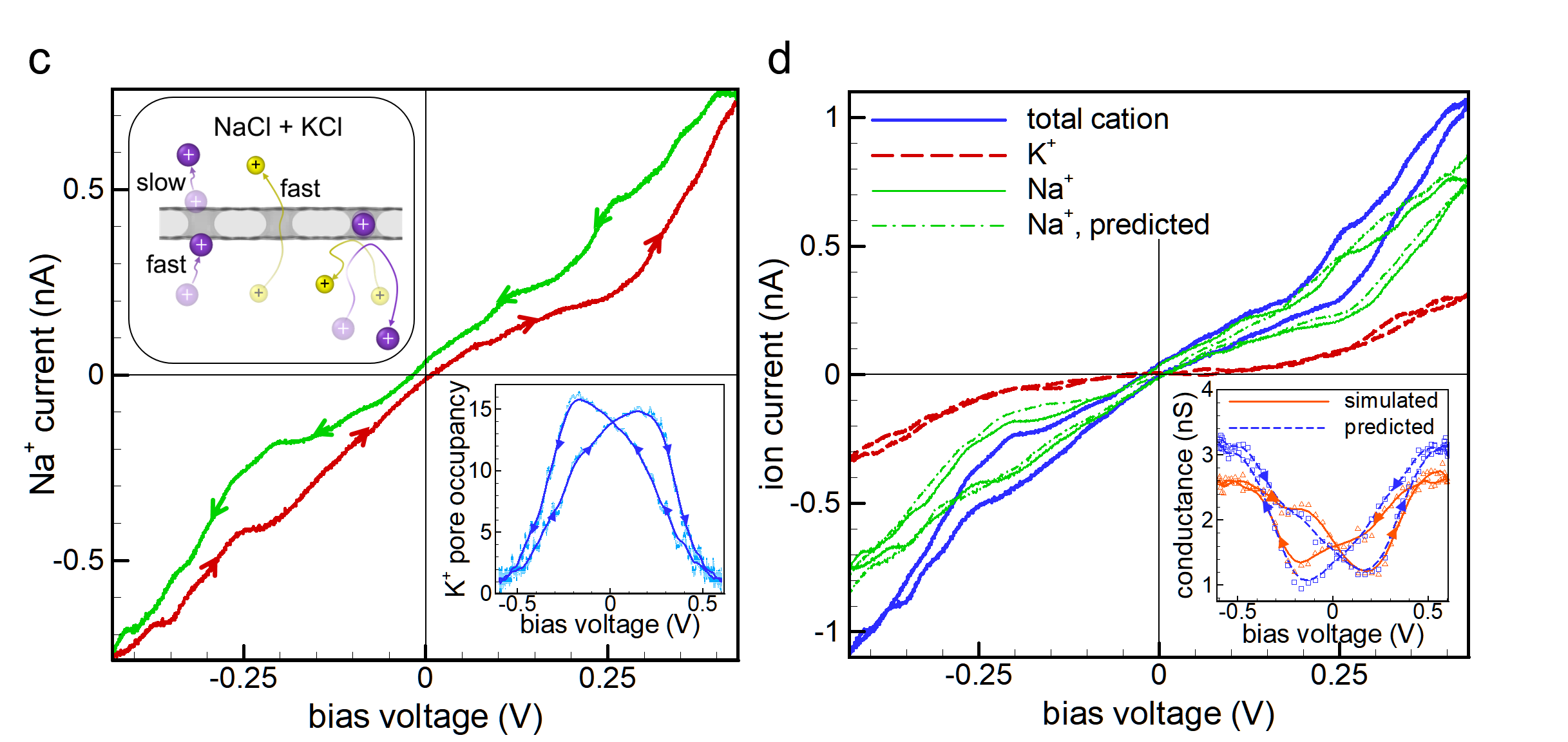}
 \vspace{0.5cm} 
 \caption{\textbf{Dynamic current-voltage response of aqueous crown-porous graphene.} Simulated system with the PMF curves for Na$^+$ and K$^+$ ions (a), Lissajous curves for Na$^+$ and K$^+$ currents in individual salt simulations of 1M NaCl and 0.25M KCl, respectively (b), Lissajous curve for Na$^+$ current obtained from a 1M NaCl + 0.25M KCl mixture (c), simulated Na$^+$ current compared with the prediction based on the pore occupancy by the K$^+$ ions, along with the total (K$^+$ and Na$^+$) cation current (d). The Lissajous curve for the K$^+$ pore occupancy is shown in the bottom right panel of (c). Inset in (d) shows the Lissajous curve for the \textit{total} simulated system conductance, compared with the analytical estimate based on the occupancy, as described in the text. The continuous curves for simulated and predicted data are visual guides obtained from smoothing the corresponding raw data points (triangles and squares, respectively).  All figures correspond to a bias field oscillation frequency of 5 MHz. Given the high density of the ionic current data presented here, the corresponding uncertainty bars are omitted for clarity. The average numerical uncertainties for the Na$^+$ and K$^+$ currents reported in this figure are 0.098 nA and 0.052 nA, respectively.} 
\label {fig:crowns}
\end{figure}
Our first example is a locally suspended graphene monolayer featuring a 4 $\times$ 5 array of \textit{18-crown-6} pores~\cite{Guo2014,Smolyanitsky2018,fang2019highly,kim2024synthesis}, immersed in an aqueous salt bath and subject to a sinusoidal external bias along the $Z$-direction, as sketched in Fig.~\ref{fig:crowns}a. The simulation details are provided in the Methods section below. A key property of these sub-nm pores is that they selectively trap aqueous K$^+$ cations, in contrast with Na$^+$, which permeate rapidly without being trapped. As discussed in detail earlier~\cite{Smolyanitsky2018,fang2019highly}, this property of crown pores is in accord with the selective binding between 18-crown-6 molecules and alkali cations~\cite{izatt1976calorimetric}, a hallmark example of selective affinity in the field of coordination chemistry. The corresponding Gibbs free energy distributions in the form of potential of mean force (PMF) curves are shown in Fig.~\ref{fig:crowns}a, indicating that in the absence of external bias K$^+$ ions encounter a significant potential well at the center of the pore, while the Na$^+$ ions do not. As discussed below, this selective affinity will prove central to the mechanism underlying memory phenomena in this case. 
Consider Fig.~\ref{fig:crowns}b, which shows the K$^+$ and Na$^+$ currents obtained from \textit{single-salt} simulations of 0.25 M KCl and 1 M NaCl, respectively. The currents are plotted as Lissajous curves, \textit{i.e.}, direct functions of the sinusoidal bias voltage $V(t)=V_0cos(\omega t)$, where in this case $V_0=h \times E_0=$ 0.6 V (h $\approx$ 6 nm, $E_0=0.1$ V/nm) and $\frac{\omega}{2 \pi}$ = 5 MHz. As shown, neither curve exhibits appreciable hysteretic behavior. Let us now combine these salts, at the same individual concentrations, into a binary mixture (0.25 M KCl + 1 M NaCl) and consider the resulting Na$^+$ current shown in Fig.~\ref{fig:crowns}c: hysteresis is now observed between the ion current branches corresponding to the rising and falling edge of the sinusoidal bias, indicative of broadly memristive transport. 
In order to understand why transport of a mixture of two salts, neither of which individually yields ion current hysteresis, exhibits memristive behavior, we must realize that crown pores conduct one ion at a time in a mutually exclusive manner. In single-salt scenarios, the cations in question permeate as sketched in the insets of Fig.~\ref{fig:crowns}b: K$^+$ ions are trapped by the pores, causing them to permeate slowly one ion at a time; in contrast, Na$^+$ ions permeate rapidly without trapping, while anions (not shown) are outright rejected due to incompatible dipole electrostatics at the pore interior~\cite{Smolyanitsky2018}. As demonstrated earlier for binary mixtures~\cite{fang2019highly}, transport occurs as a mostly unidirectional competition between Na$^+$ and K$^+$ ions. Recall that crown pores have a high selective affinity for the K$^+$ ions. Therefore, Na$^+$ cations are statistically expected to \textit{permeate through pores unoccupied by the trapped K$^+$ ions}, as sketched in the upper-left inset of Fig.~\ref{fig:crowns}c. In contrast, K$^+$ transport is not hampered by the presence of Na$^+$, because the latter permeate rapidly without being trapped. A reasonably detailed treatment based on the modified Langmuir model~\cite{Langmuir1918} is possible for this competition, as presented in the supplementary section S3 of our earlier work~\cite{fang2019highly}. For the purpose of this discussion, a more tacit argument is provided. Depending on the bias, the number of pores clogged by the trapped K$^+$ ions is a dynamic function that naturally carries delay from the finite times it takes the K$^+$ ions to \textit{cumulatively} occupy or leave the pores, as shown for sinusoidal bias in the bottom-right inset of Fig.~\ref{fig:crowns}c. Given that the time taken by the K$^+$ ions to populate the pore array is mainly determined by a diffusive process~\cite{accres}, the corresponding time delay associated with memristive response is expected to be tunable by the K$^+$ concentration. The latter could be adjusted to achieve memristive response at bias frequencies significantly lower than those considered here. It should then be clear that we selected a relatively low KCl concentration of 0.25 M (compared with 1 M NaCl) in order to achieve an observable delay and to reduce the K$^+$ contribution to the overall cation transport, all while maintaining robust data collection within microseconds of total simulated time.
Neglecting the ion-ion knock-on phenomena, the pore occupancy by the pore-clogging K$^+$ ions is thus the system state that governs Na$^+$ transport in a mixture. Let us denote the occupancy $N_{K^+}(t)$ and point out a few noteworthy aspects. First, K$^+$ permeation, both as single salt or in a mixture involving Na$^+$, does not depend on $N_{K^+}(t)$~\cite{fang2019highly}, because the charge carriers and the pore-blocking ions are identical and thus essentially all pores are available for K$^+$ permeation.  This is consistent with nearly the same \textit{non-hysteretic} K$^+$ current data in Figs.~\ref{fig:crowns}b and~\ref{fig:crowns}d. Second, $N_{K^+}(t)$ is likely to be directly observable in the form of electrical potential measured at the membrane, as pointed out earlier for DC-bias~\cite{Smolyanitsky2018}. This measurement of electrical potential is expected to be significantly less affected by the bandwidth limitations associated with, for example, ion current measurements. Finally, $N_{K^+}(t)$ can be used as an elementary coupling term to predict the Na$^+$ current for the NaCl+KCl mixture. If the total number of pores in the system is $N_0=20$, only those \textit{unoccupied} by K$^+$ are available for Na$^+$ transport. The Na$^+$ current for a mixture is then estimated as $I^{mix}_{Na^+}(t) = I_{Na^+}(t) \left( 1 - \frac{N_{K^+}(t)}{N_0} \right)$, where $I_{Na^+}(t)$ is the \textit{non-hysteretic} single-salt Na$^+$ current response from Fig.~\ref{fig:crowns}b. The comparison between this estimate and that simulated in an actual mixture (data also shown in Fig.~\ref{fig:crowns}c) is shown in Fig.~\ref{fig:crowns}d. An overall reasonable agreement between the two Lissajous curves suggests that the observed dynamic mixture sieving is an example of a system where the current-voltage response is an illustratively simple linear function of the system state. For further qualitative comparison, see the inset of Fig.~\ref{fig:crowns}d, where the total conductance $G(t)=I(t)/V(t)$ (including both K$^+$ and Na$^+$ contributions) simulated in the mixture is compared with the analytical estimate given by $I^{tot}(t) = I^{mix}_{Na^+}(t) + I_{K^+}(t)$, where $I^{mix}_{Na^+}(t)$ is as defined above and $I_{K^+}(t)$ is the non-hysteretic K$^+$ contribution from the single-salt data in Fig.~\ref{fig:crowns}b. Both curves are shown to qualitatively inherit the time dependence of K$^+$ occupancy as a function of time (see bottom right inset of Fig.~\ref{fig:crowns}c). 

\begin{figure}[ht!]
\centering
\includegraphics[width=0.95\linewidth]{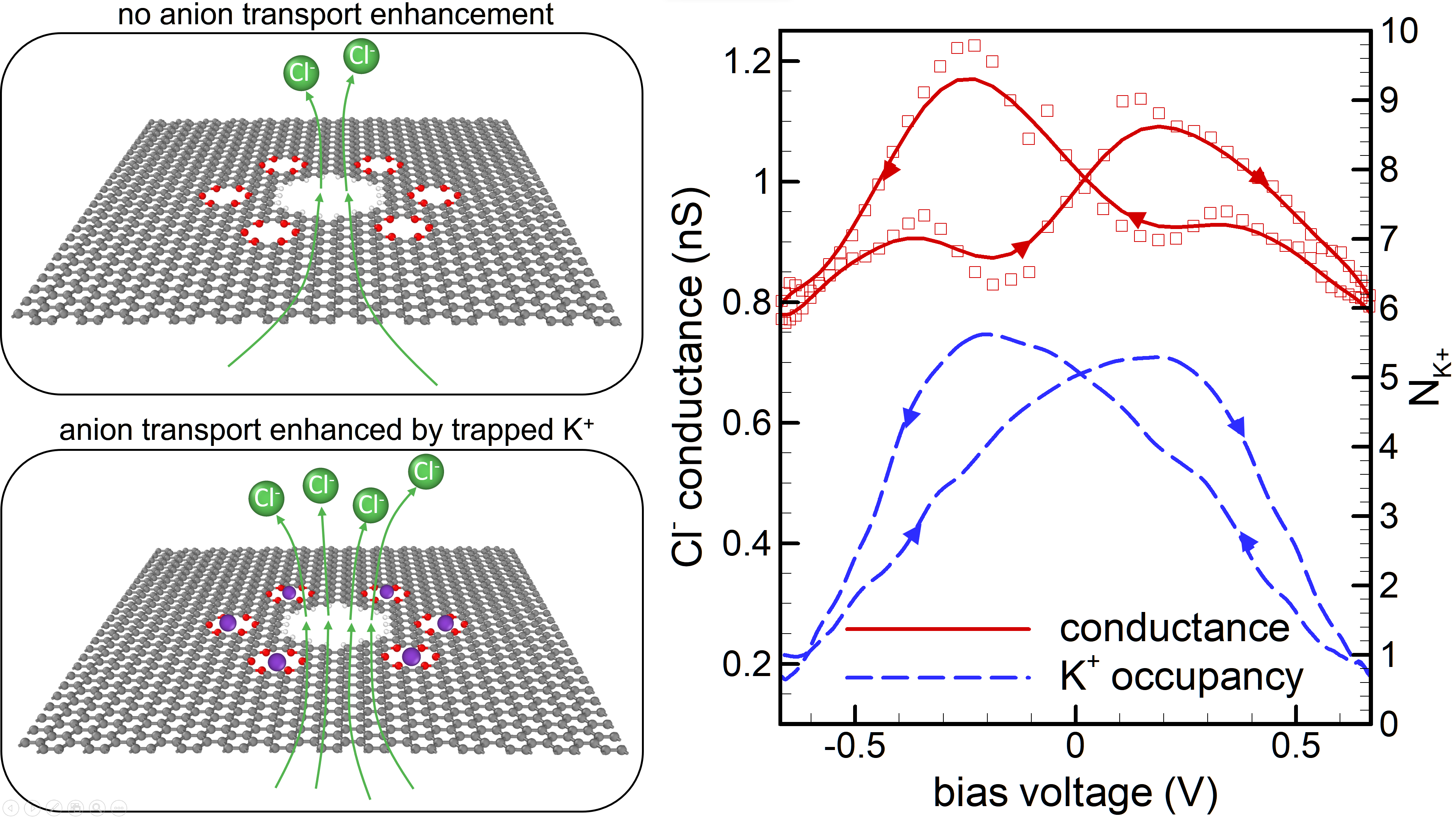}
\caption{\textbf{Anion transport enhanced by trapped K$^+$.} The distinct anion transport states for the simulated porous membrane are shown on the left. Anion conductance $G_{Cl^-}(t) = I_{Cl^-}(t)/V(t)$ and K$^+$ occupancy are plotted on the right. The continuous Lissajous curve for $G_{Cl^-}(t)$ is a visual guide obtained from smoothing the corresponding raw data points (squares). The data on the right was simulated in a box with $h$ = 9 nm and $E_0=0.075$ V/nm at $\frac{\omega}{2 \pi}=$ 5 MHz. The estimated bias voltage amplitude $V_0=h \times E_0=$ 0.675 V. The average numerical uncertainty for the Cl$^-$ ionic currents underlying the data in this figure is 0.045 nA.}
\label {fig:coupled}
\end{figure}

The state-transport coupling through direct blockage of conductive paths by trapped ions is one example. A less direct coupling is demonstrated in Fig.~\ref{fig:coupled}, where the system is similar to the one above, except the graphene membrane features a relatively wide (d $\approx$ 1.5 nm) pore, closely surrounded by six crown-like pores. Here, K$^+$ trapping results in a ring of temporarily immobile charge, which at sufficiently low ion concentrations is expected to modulate anion transport~\cite{PhysRevLett.93.035901,Smolyanitsky2009,joshi2010} by attracting counterions (Cl$^-$) toward the large pore. This field-induced coupling is directly observable here, because anions have no affinity for the crown pores, permeating only through the wide pore. No cation mixture is required in this case and we use KCl at a concentration such that the corresponding Debye screening length is comparable to the radius of the wide pore (0.15 M). Although less pronounced than in the case above, memristive anion transport is indeed revealed in the right panel of Fig.~\ref{fig:coupled}, similar to that in the inset of Fig.~\ref{fig:crowns}d, except here the hysteretic accumulation of trapped K$^+$ enhances transport of another ionic species instead of hampering it. 

\begin{figure}[ht!]
\centering
\includegraphics[width=1.0\linewidth]{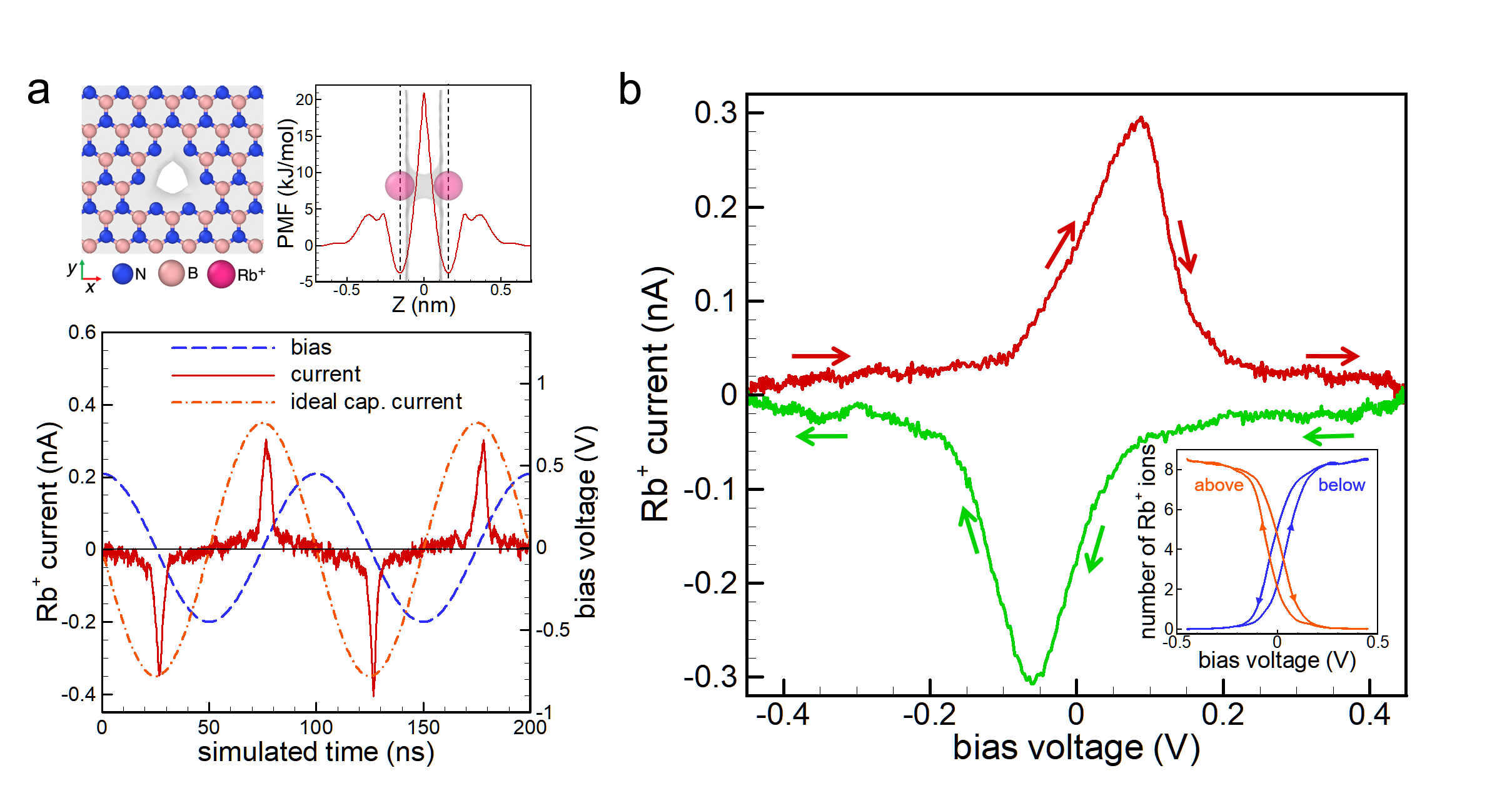} 
  \vspace{-0.5cm} 
 \includegraphics[width=1.0\linewidth]{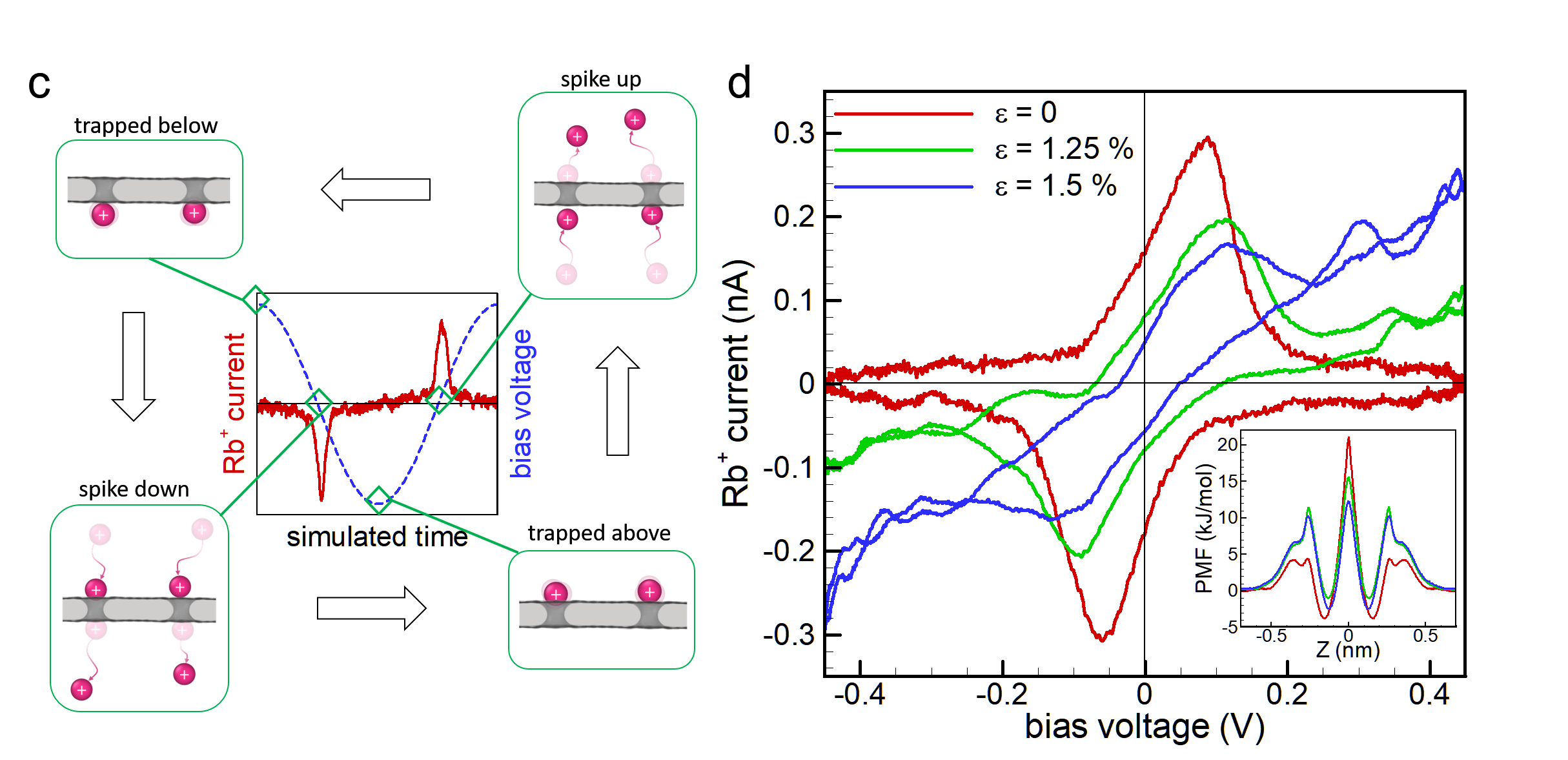}
\vspace{0.5cm} 
 \caption{\textbf{Spiking capacitive Rb$^+$ currents through sub-nm-porous hBN}. Pore structure, the corresponding PMF curve for Rb$^+$ ions, and the simulated ionic current as a function of simulated time (a), along with the the corresponding Lissajous curve (b); a sketch of the system state dynamics (c), Lissajous curves for Rb$^+$ currents through pores with the corresponding PMF curves shown in the inset (d). The inset in (b) shows the Lissajous curves of the ion occupancy above and below the membrane's geometric mid-line. All figures correspond to a bias field oscillation frequency of 10 MHz. The average numerical uncertainty for the ionic currents reported in this figure is 0.025 nA.}
\label {fig:hbn}
\end{figure}

Our second example is similar to those above, except the membrane is monolayer hBN featuring a 3 $\times$ 3 array of nitrogen-terminated triangular pores, one of which is shown at the top of Fig.~\ref{fig:hbn}a. For these sub-nm pores, the dipolar pore edge with partial negative charges carried by the nitrogen atoms at the perimeter~\cite{Liu2017} generally yields crown-like properties which include cation trapping and broad anion rejection, thus making the phenomena described above generally expected here as well. On this occasion, however, we wish to explore a pore impermeable to a selected ionic species, resulting in a qualitatively different scenario. Our salt choice is 1 M RbCl and the corresponding PMF curve for the Rb$^+$ ion is provided in Fig.~\ref{fig:hbn}a. In addition to being impermeable due to a large peak at the pore's geometric mid-line, the pore features two relatively weak binding sites $\approx 0.16$ nm above and below the mid-line. Note that ion-ion Coulomb repulsion causes the two bound states to be mutually exclusive, \textit{i.e.}, only one ion at a time can bind on either side. 

The Rb$^+$ current as a function of time in response to a sinusoidal bias with $V_0=$ 0.45 V and $\frac{\omega}{2 \pi}=$ 10 MHz is shown at the bottom of Fig.~\ref{fig:hbn}a, alongside $V(t)$ and the ideal capacitive current ($\propto\frac{dV}{dt}$) provided as references. In contrast with the gradual discharge of an ideal capacitor, we see bidirectional $\approx$15 ns-wide current spikes near $V=0$ crossings, all attributed to rapid release of ions, depending on the bias polarity in relation to the current charging state. The corresponding Lissajous curve is shown in Fig.~\ref{fig:hbn}b. The spiking dynamics here is due to the presence of a chemical barrier built into this capacitor and a simple "state diagram" in Fig.~\ref{fig:hbn}c is provided to describe the switching cycle and explain the presence of elementary directional memory. Overall, this system can be viewed as a simple spiking NOT-gate with polarity awareness: zero bias voltage (input) results in a current (output) up-spike if the previous input was negative and \textit{vice versa}. We note that the charge/discharge curves (in the form of pore occupancies) ``above" and ``below" shown in the inset of Fig.~\ref{fig:hbn}b are indicative of an accompanying memcapacitive effect, resulting from time delays caused by the mechanisms similar to those discussed above. 
The ``firing" dynamics of the ionic currents should be tunable in terms of the peak height and width/phase, as determined by the total pore count and the barrier shape, respectively. We also note that reduced binding (while maintaining pore impermeability) would bring the peaks of the two Lissajous branches in Fig.~\ref{fig:hbn}b closer to zero-voltage and also widen them toward semi-circles, essentially approaching the dynamic response of an ideal capacitor. While testing these behaviors would be beyond the scope of this work, here we can briefly explore a change in transport response as we tune the ion-pore interactions by applying isotropic strain to the hBN membrane. The Lissajous curves for the Rb$^+$ currents, along with the corresponding PMF data, are shown in Fig.~\ref{fig:hbn}d. It is clear that by applying membrane strain and dilating the pores, the central barrier is reduced, causing ion permeation and thus a transition from capacitive to RC-circuit behavior, which at even higher strains we expect to approach ohmic response. Once again, the simplicity here is illustrative, suggesting nanofluidic "gates" with dynamic transport response tuned by strain or an auxiliary electrostatic bias, possibly dynamically. 

To summarize, we have demonstrated memristive effects and spiking behavior of dynamically biased aqueous ion transport through 2D materials featuring arrays of crown-like sub-nm pores. The mechanisms are shown to include competitive sieving of ion mixtures, resulting in a coupling between the time-delayed state of the system and its transport properties, as well as capacitive charging and discharging in the presence of built-in chemical barriers. The phenomena described above are highly illustrative and suggest that nanofluidic systems based on subnanoporous 2D materials may be an intriguing choice for achieving analog-digital hybrids usable in artificial neural networks, especially if aimed at dynamics in the range of tens to hundreds of kHz to MHz. By focusing on the mechanisms rather than specific applications, our hope is to elucidate the physics of realistically observable dynamic effects in nanofluidic ion transport and further stimulate ongoing experimental efforts. First and foremost, this includes fabrication of predictable, chemically stable pore structures with various degrees of affinity to aqueous solutes.

\section{Methods}
All MD simulations were performed using GPU-accelerated GROMACS v. 2023.2 within the OPLS-AA~\cite{Jorgensen1996} framework. Each simulation of ion transport was carried out in a nearly cubic box with a side of $\approx$ 6 nm, periodic in $XYZ$ and containing a porous monolayer of graphene or hBN, TIP4P~\cite{tip4p} water, and dissociated salts at concentrations stated in the main text. The membranes were kept in place by harmonic restraints applied at the perimeter. Previously established models for crown-porous graphene~\cite{Smolyanitsky2018, fang2019highly} and hBN~\cite{hbnff_GovindRajan2018} were used within the OPLS-AA forcefield. The partial charges of nitrogen atoms lining the triangular pore edges in hBN were set to $2/3$ of those for bulk nitrogen atoms, ensuring charge neutrality of the pore structures. Electrostatic interactions were resolved using the particle-particle—particle-mesh scheme with a short-range interaction cut-off radius of 1.0 nm and 1.2 nm for the simulations involving graphene and hBN, respectively. 
Prior to the production simulations, all systems underwent static energy minimization, 5 ns of semi-isotropic (constant in-plane cell dimensions) NPT relaxation at T = 300 K and P = 0.1 MPa with a time-step of 1 fs. Each production simulation was performed in the NVT ensemble with a time step of 2 fs under sinusoidal external field $E(t)=E_0cos(\omega t)$, preceded by a 10-ns-long pre-relaxation at a constant field $E_0$ to reduce any spurious oscillations from the initial impulse ($E_{t=0}=E_0$). The simulated times were set to ensure 8-10 full periods of external field variation. For example, for $\frac{\omega}{2\pi}$ = 5 MHz, the total simulated time was 2 $\mu s$, corresponding to ten 200-ns-long periods. Unless stated otherwise, $E_0$ was set to 0.075 V/nm, roughly corresponding to an effective maximum voltage of $E_0 \times h$ = 0.45 V (h $\approx$ 6 nm is the box height in the $Z$-direction). The $E_0$ values were selected to ensure that the crown pores are depopulated by the K$^+$ ions at maximum bias~\cite{Smolyanitsky2018}.
All ion current data was obtained from numerical differentiation of the cumulative ionic fluxes, performed using finite differences at the 8$^{th}$ order of accuracy. Numerical differentiation is a noise-amplifying procedure and therefore prior to differentiation all flux data had to be filtered without introducing purely numerical memory artifacts, which was achieved through careful use of bidirectional filtering (see details in section S1 of the Supplementary Material). The raw flux and occupancy data was output every 10 ps, corresponding to a maximum resolvable frequency of 50 GHz. Given that in this work the external bias oscillated at 5-10 MHz, low-pass cutoffs were typically set to the order of 200 MHz to provide ample bandwidth for capturing the dynamics of interest. The ion currents used for constructing the Lissajous curves were obtained from averaging between all simulated periods of external field oscillation and the statements regarding uncertainties refer to the corresponding standard deviations. For further detail on data processing used in this work, refer to the section S1 of Supplementary Material. All PMF calculations were carried out similarly to our previous work~\cite{Smolyanitsky2018,fang2019highly,fang2019mos2}, utilizing the Weighted Histogram Analysis Method~\cite{wham_Hub2010} applied to a total of 60 0.05-nm-spaced ionic configurations relative to the pore location along the $Z$-direction. The umbrella sampling of each configuration of the target ion was performed for 10 ns.

\section*{Acknowledgments}
This work was supported in part by the NSF award \textnumero 2110924. The authors are grateful to Frances Allen and Dana Byrne for illuminating discussions.
\section*{Author contributions}
Y.N. carried out simulations, processed and analyzed data, and wrote the first version of the manuscript. A.S. conceived the central concepts, designed and carried out simulations, and supervised the project.  Both authors discussed the results, wrote the manuscript, and contributed to revisions.

\bibliography{sn-bibliography} 

\end{document}


\beginsupplement
\title[Article Title]{\textbf{\textit{Supplementary material for:} Memristive response and capacitive spiking in the aqueous ion transport through 2D nanopore arrays}}

\author[1,2]{\fnm{Yechan} \sur{Noh}}

\author[1]{\fnm{Alex} \sur{Smolyanitsky}}

\affil[1]{\orgdiv{Applied Chemicals and Materials Division}, \orgname{National Institute of Standards and Technology}, \orgaddress{\city{Boulder}, \postcode{80305}, \state{Colorado}, \country{United States}}}

\affil[2]{\orgdiv{Department of Materials Science and Engineering}, \orgname{University of California, Berkeley}, \orgaddress{\city{Berkeley}, \postcode{94720}, \state{California}, \country{United States}}}
\newcommand*\mycommand[1]{\texttt{\emph{#1}}}
\maketitle
\section{Data processing}
\begin{figure}[ht!]
\centering
\includegraphics[width=1.0\linewidth]{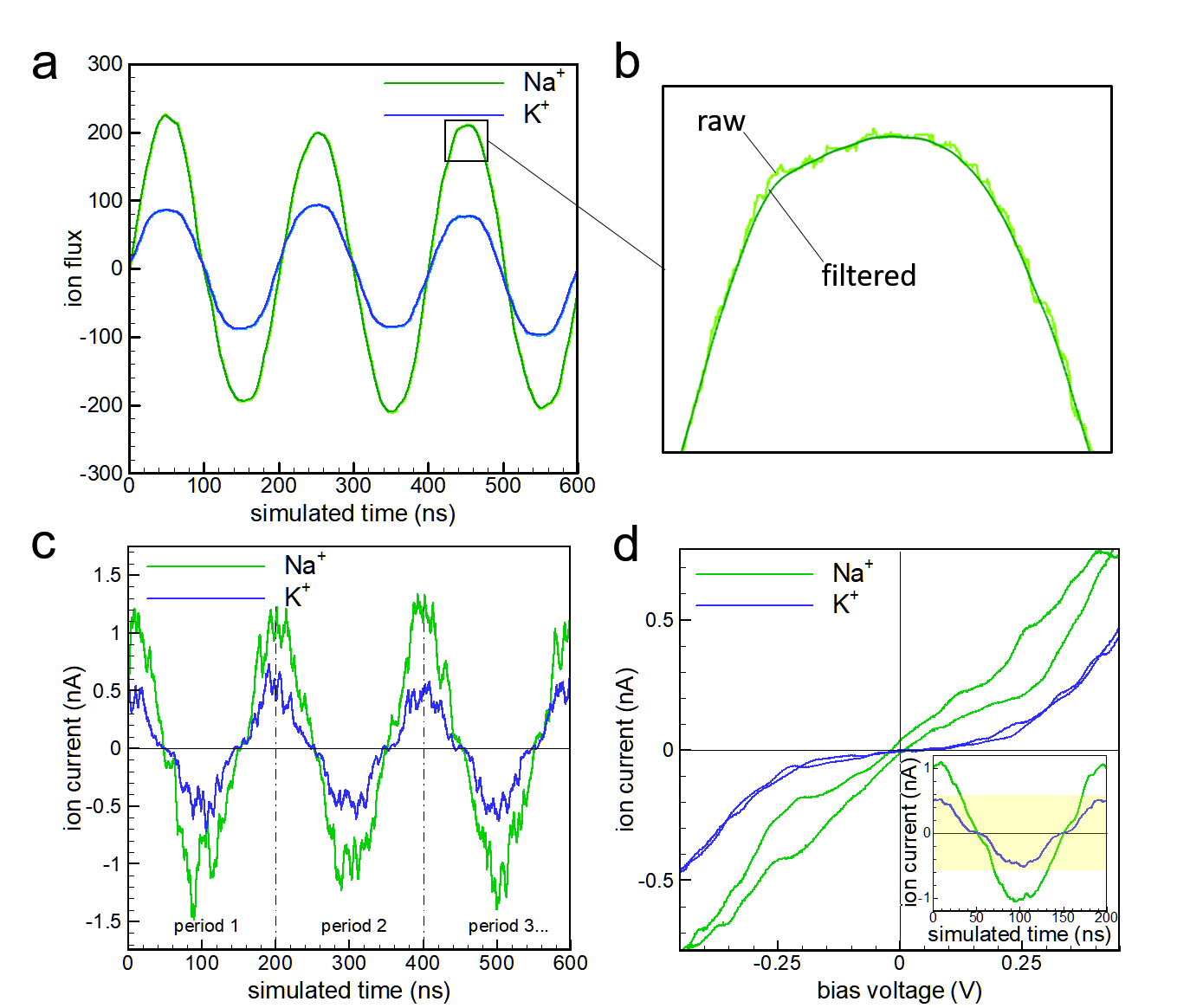}
 \caption{Data processing scheme used in this work: ion flux filtering (a,b), followed by numerical differentiation of the filtered flux data to obtain the currents (c) and calculating an averaged (over multiple cycles) single period in the time domain, as well as in the form of a Lissajous curve (d). The inset in (d) is shows the averaged periods of ion current oscillation with the yellow box outlining the Na$^+$ current range of interest reported in the (d) panel. The raw data used to obtain Fig. 1 in the main text is used here as an example. The complete dataset for this simulation was 2-$\mu$s-long, corresponding to a total of ten 200-ns-long periods; for clarity, only 600 ns are shown.} 
\label {fig:dataproc}
\end{figure}

\begin{figure}[ht!]
\centering
\includegraphics[width=0.75\linewidth]{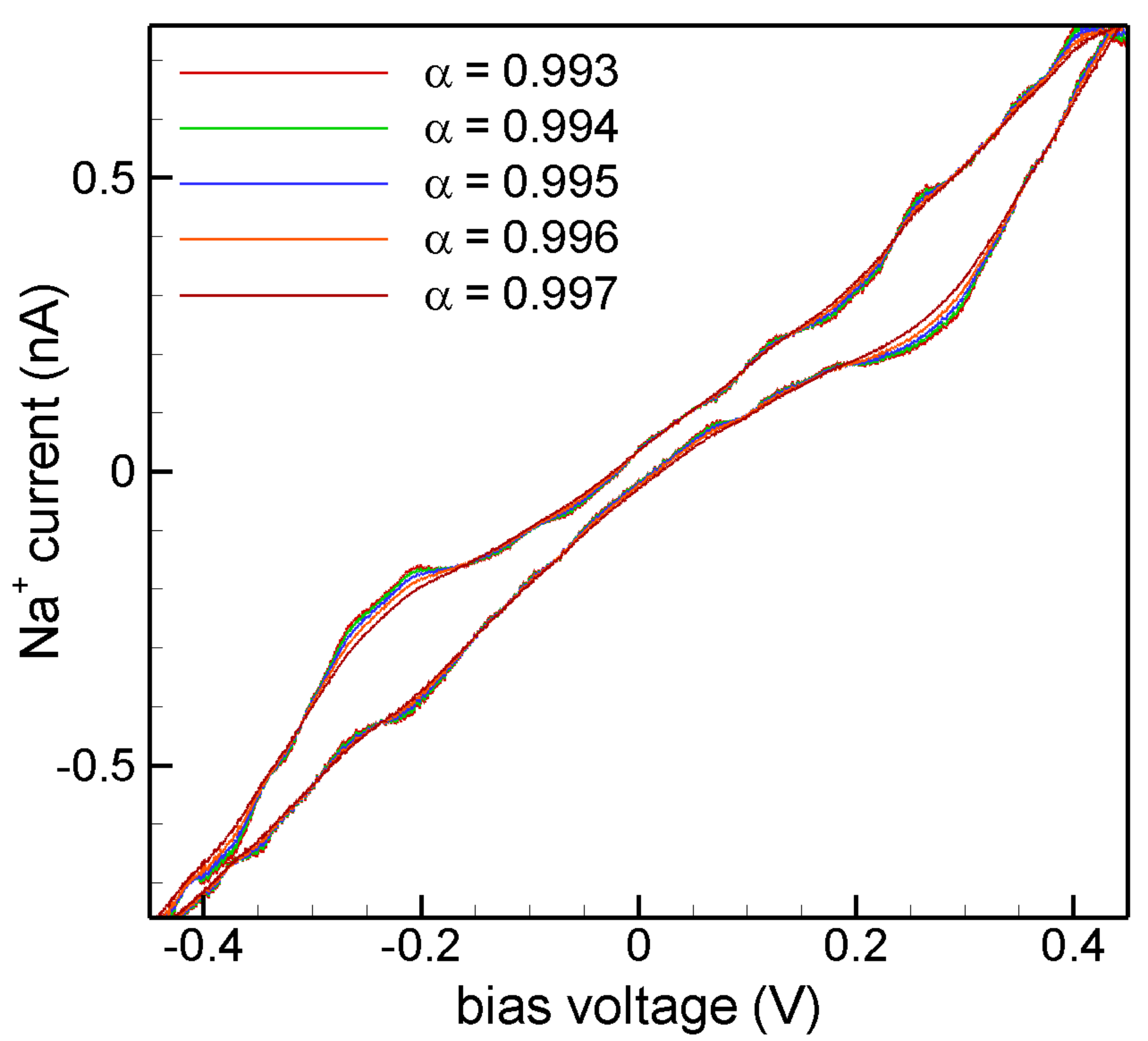}
 \caption{Low-pass cut-off parameter sweep for the Lissajous curve corresponding to Na$^+$ transport.} 
\label {fig:alphasweep}
\end{figure}

A three-step data processing procedure was used to obtain the transport data presented in the main text, as shown in Fig.~\ref{fig:dataproc}.  The first step is low-pass filtering of the raw ionic flux data (Fig.~\ref{fig:dataproc}ab), followed by numerical differentiation to obtain the corresponding currents (Fig.~\ref{fig:dataproc}c), and finally displaying the current in the form of a Lissajous IV curve (Fig.~\ref{fig:dataproc}d). Similarly to the previous experimental works reporting memristive devices cited in the main text, the Lissajous curve was based on a representative average period calculated from multiple periods in Fig.~\ref{fig:dataproc}c. This averaged data is shown in the inset of Fig.~\ref{fig:dataproc}d.  
\newline
First, low-pass filtering was applied to the raw flux data to remove high-frequency noise. This processing step preceded numerical differentiation (see below), a noise-amplifying calculation, making low-pass filtering critical in reducing spurious thermal noise while keeping the useful data within the frequency range of interest (5-10 MHz) reasonably intact. The low-pass cutoff was set as follows. For electrostatic biases in this work (in the 5-10 MHz range), the effective bandwidth of the filter was set to the order of 200 MHz, as stated in the Methods section of the main text. We used several filters (third-order Chebyshev finite impulse response (FIR) filter, as well as an infinite impulse response (IIR) filter) independently to ensure nearly identical final results. As an example, the IIR filter used in this work was given by $y(n) = \alpha y(n-1) + \frac{1-\alpha}{N}\sum_{k=0}^{k=N-1} x(n-k)$, where $x(n)$, $y(n)$, and $N=6$ are the raw data, filtered data, and the running average input width, respectively, while $1 -\alpha$ directly set the cutoff. For instance, our raw flux data points were spaced by $\tau$=10 ps, which corresponds to an effective Nyquist frequency of $f_0 = \frac{1}{2\tau} = $ 50 GHz. To achieve a desired effective bandwidth of $f_c$ = 200 MHz, the corresponding filter setting is $\alpha = 1 - \frac{f_c}{f_0}$ = 0.996. We found the Lissajous curves to be similar between $\alpha$ = 0.993 and $\alpha$ = 0.997, as shown in Fig.~\ref{fig:alphasweep}. One must keep in mind that the presented parameterization is specific to the particular filter described above and various filters with appropriate low-pass cutoffs can be used to achieve similar results.
\newline 
It is critically important to note that all low-pass filtering is guaranteed to introduce a purely numerical phase shift between the raw and filtered data. To eliminate this shift, flux data was filtered bidirectionally, \textit{i.e.}, the filter described above was applied twice: first, starting at the beginning of the raw flux dataset and moving forward in time and second, starting at the end of the dataset and moving backward in time. The production filtered flux data (in Fig.~\ref{fig:dataproc}ab) was calculated as an average between the forward- and backward-filtered sequences, ensuring no phase shifting at the numerical level (see Fig.~\ref{fig:dataproc}b, for example). Alternatively, one can employ Python's \textit{filtfilt} procedure~\cite{filtfilt} to achieve a similar shift-free filtering outcome. 
\newline
Once filtered flux data was obtained, a 9-point (accurate to 8$^{th}$ order) finite difference~\cite{findif} was used to calculate the ion currents as $q\frac{df}{dt}$, where $q$ is the ion charge and $f(t)$ is filtered flux (Fig.~\ref{fig:dataproc}a). The corresponding raw currents (Fig. ~\ref{fig:dataproc}c) were then used to obtain a representative single-period (inset of Fig.~\ref{fig:dataproc}d) as an average between all periods. This averaged single-period data (inset of Fig.~\ref{fig:dataproc}d) was finally used to generate the corresponding Lissajous curves (Fig.~\ref{fig:dataproc}d), in which the simulated time along the abscissa is merely replaced by the bias $V(t)=V_0cos(\omega t)$. 
\newline
The numerical uncertainties reported in the main text are not error bar equivalents for the specific data of interest. They are the per-point standard deviations between the entire averaged period and the ten periods used to obtain it. Such uncertainties represent the all-encompassing data variability, \textit{including the noisiest regions near the sinusoidal extrema} (see Fig.~\ref{fig:dataproc}c), which are not part of the Lissajous curves of interest in Fig.~\ref{fig:dataproc}d. As expected, this uncertainty depends on the filter cut-off: for the curves in Fig.~\ref{fig:alphasweep} it ranges between 0.067 nA and 0.11 nA at $\alpha$ = 0.997 (smaller bandwidth) and $\alpha$ = 0.993 (larger bandwidth), respectively.